\documentclass[12pt]{article}
\pdfoutput=1
\usepackage{appendix}
\usepackage{bbm}
\usepackage{color}
\usepackage{authblk}
\usepackage{latexsym}
\usepackage{epsfig,amssymb,euscript, mathrsfs,cite}
\usepackage{amsmath}
\usepackage{mathrsfs} 
\usepackage{enumerate}
\definecolor{MyDarkBlue}{rgb}{0.15,0.15,0.45}
\usepackage[linktocpage=true]{hyperref}
\hypersetup{
colorlinks=true,
citecolor=MyDarkBlue,
linkcolor=MyDarkBlue,
urlcolor=MyDarkBlue,
pdfauthor={},
pdftitle={},
pdfsubject={hep-th}
}

\newsavebox{\ns}
\newsavebox{\dbrane}
\newsavebox{\dbshort}

\def\be{\begin{equation}}
\def\ee{\end{equation}}
\def\bea{\begin{eqnarray}}
\def\eea{\end{eqnarray}}

\def\eq#1 { \begin{equation} #1 \end{equation} }

\newlength{\sswidth}

\numberwithin{equation}{section}       

\usepackage{amsmath}
\usepackage{amsthm}
\usepackage{thmtools}
\usepackage{amsfonts}
\usepackage{amssymb}

\newtheorem{conjecture}{Conjecture}

\newcommand*\qm[1][!]{\ifx#1! \mu \else \mu_{#1} \fi} 

\newcommand*\hi[1][!]{\ifx#1! \gamma \else \gamma_{ #1 } \fi} 
\newcommand*\HS[1][!]{\ifx#1! \Omega \else \Omega_{#1} \fi} 
\newcommand*\Eset[1]{\mathfrak{#1}} 
\newcommand*\EA[1][!]{\ifx#1! \Eset{A} \else \Eset{A}_{#1} \fi} 
\newcommand*\QuasiSystem[1][!]{\ifx#1! (\HS,\EA,f) \else (\HS[#1],\EA[#1],f_{#1}) \fi} 
\newcommand*\System[1][!]{\ifx#1! (\HS,\EA,D) \else (\HS[#1],\EA[#1],D_{#1}) \fi} 

\newcommand*\Sys[1][!]{\ifx#1! \Psi \else \Psi_{#1} \fi} 

\begin{document}

\begin{titlepage}

\vfill



\begin{center}
   \baselineskip=16pt
   {\Large\bf Boundary contributions in the causal set action}
  \vskip 1.5cm
Fay Dowker${}^{a,b}$\\
     \vskip .6cm
            \begin{small}
      \textit{
      		${}^{a}${Blackett Laboratory, Imperial College, Prince Consort Road, London, SW7 2AZ, UK}\\ \vspace{5pt}
${}^{b}${Perimeter Institute, 31 Caroline Street North, Waterloo ON, N2L 2Y5, Canada}\\
 \vspace{5pt}
 email: f.dowker@imperial.ac.uk
              }
              \end{small}                       \end{center}
\vskip 2cm
\begin{center}
\textbf{Abstract}
\end{center}
\begin{quote}
Evidence is provided for a conjecture that, in the continuum limit,  the
mean of the causal set action of a causal 
set sprinkled into a globally hyperbolic Lorentzian spacetime, $\mathcal{M}$, of finite volume equals the
Einstein Hilbert action of $\mathcal{M}$ plus the volume of the
co-dimension 2 intersection of the future boundary with the past boundary. We give the heuristic argument
for this conjecture and analyse
some examples in 2 dimensions and 
one example in 4 dimensions. \end{quote}

\vfill

\end{titlepage}

\tableofcontents

\newpage

\section{The causal set action}\label{sec:intro}

The Benincasa-Dowker-Glaser causal set action \cite{Benincasa:2010ac,Benincasa:thesis,
Dowker:2013vba,Glaser:2013xha} is a family of actions, $S^{(d)}_{BDG}(\mathcal{C})$ for a finite causal set, $\{\mathcal{C}, \preceq\}$, one action for each natural number $d>1$. 
\begin{equation}\label{bdgaction}
\frac{1}{\hbar}\mathcal{S}_{BDG}^{(d)}(\mathcal{C}) = \zeta_d \bigg( N + \frac{\beta_d}{\alpha_d}
 \sum_{i = 1}^{n_d} C_i ^{(d)} N_i \bigg)\,,
 \end{equation}
where $\zeta_d := - \alpha_d (\frac{l}{l_p})^{{d-2}}$,  and $\alpha_d$ and $\beta_d$ are $d$-dependent constants of order 1. $l_p^{d-2} =8 \pi G\hbar$ is the $d$-dimensional Planck length and $l$ is the fundamental length scale of causal set theory so the ratio $\frac{l}{l_p}$ is expected to be 
a dimensionless number of order 1.
$N_i$ is the number of 
inclusive order intervals of cardinality $i+1$ in $\mathcal{C}$, 
where the order interval $I(a,b)$ between two causal set elements $a$ and $b$ such that $a\prec b$ is given by $I(a,b) : = \{ c \in \mathcal{C}\, |\, a \preceq c \preceq b \}$. $n_d := \left \lfloor{\frac{d+4}{2} }\right \rfloor$ so 
for $d=2,3$ there are 3 terms in the sum, for $d=4,5$ there are 4 terms \textit{etc.} 
$C^{(d)}_1$ is fixed to be
equal to $1$ and the other $C_i^{(d)}$ are rational constants 
of alternating sign. The values of the constants $\alpha_d$, $\beta_d$, and $C_i^{(d)}$ for all $d$
are given in \cite{Glaser:2013xha}.

For each globally hyperbolic Lorentzian spacetime $\mathcal{M}$ of dimension $d$ and finite volume, the Poisson process of sprinkling at density $\rho := l^{-d}$ and the causal set action $S^{(d)}$ gives rise to a random variable $\mathbf{S}_\rho(M)$ that equals the action evaluated on the random causal 
set that is the outcome of the sprinkling process.\footnote{Out of the family of actions, it is the one where $d$ equals the dimension of $\mathcal{M}$ that is used to construct the random discrete action of $\mathcal{M}$. For that reason, we drop the superscript $d$ on $\mathbf{S}_\rho(M)$ as it is implied by the dimension of $\mathcal{M}$.}  We call this random variable the random discrete action of $\mathcal{M}$ at density $\rho$. It is conjectured that in the continuum limit of $l \rightarrow 0$ or $\rho \rightarrow \infty$, the 
expected value of the random discrete action,  $ \langle \mathbf{S}_\rho(M) \rangle$,
 tends to ($\hbar$ times) the Einstein-Hilbert action plus certain boundary contributions \cite{Benincasa:thesis}.\footnote{The random discrete action can be defined for spacetimes that satisfy weaker causality conditions than global hyperbolicity. For example, the random discrete action of the 2 dimensional trousers spacetime 
 is studied in \cite{Benincasa:2010as}.}
 
More precisely, let $\mathcal{M}$ be globally hyperbolic and of finite volume. Then, the boundary of (the closure of) $\mathcal{M}$ is achronal: no two points of the boundary are timelike related.
 The boundary of $\mathcal{M}$ is the union of 
 $\Sigma_-$ and $\Sigma_+$, the past and future boundaries respectively. $\Sigma_{+}$ ($\Sigma_{-}$) is defined to be the set of points at which future (past) going  timelike curves leave the closure of $\mathcal{M}$. The hypersurface $\Sigma_{\pm}$ can be null (for example a causal interval), 
 spacelike (for example a slab of an Einstein static cylinder) or both (for example a ``truncated'' causal interval with its top sliced off).

\begin{conjecture}\label{conjecture}
Page 44 of \cite{Benincasa:thesis}:
\begin{equation}
\lim_{\rho\rightarrow \infty} \frac{1}{\hbar} \langle \mathbf{S}_\rho(M) \rangle 
= \frac{1}{l_p^{d-2}}\int_{\mathcal{M}} d^dx \sqrt{-g} \frac{R}{2} + \frac{1}{l_p^{d-2}}{\textrm{Vol}}_{d-2}(J)\,,
\end{equation}
where $J : = \Sigma_- \cap \Sigma_+$, which we will refer to as the joint, and ${\textrm{Vol}}_{d-2}(J)$ is
its volume. 
\end{conjecture}

There is some evidence for the conjecture in the literature for the case of flat spacetime: it holds  for flat causal intervals in all dimensions \cite{Benincasa:2010as,Buck:2015oaa} and for a null triangle and for a cylinder spacetime  in 2 dimensions \cite{Benincasa:2010as}. 

To understand the conjecture, recall where the action comes from. 
The family of actions action arose from the discovery of  scalar d'Alembertian analogues on causal sets, starting with $d=2$ \cite{Sorkin:2007qi} and $d=4$ \cite{Benincasa:2010ac} and then for all $d$ \cite{Dowker:2013vba,Glaser:2013xha}. 
For a scalar field on $\mathcal{C}$,  $\phi:\mathcal{C} \rightarrow \mathbb{R}$, for each $d>0$, there is a retarded d'Alembertian operator $B^{(d)}$: 
\begin{equation}\label{Bretard}
B^{(d)}\phi(a)=\frac{1}{l^2}\left( \alpha_d \phi(a) +\beta_d \sum\limits_{i=1}^{n_d}  C^{(d)}_i \sum\limits_{b \in L_{i}}\phi(b)\right) \, ,
\end{equation} 
where 
$a\in \mathcal{C}$ and the sums are over levels $L_i : = \{ c \in \mathcal{C}\, |\, c \prec a\ \ \textrm{and} \ \ 
| \, I(a,c) \, | = i +1 \}$. So, for example the first level, $L_1$, is the set of elements that precede $a$ and are linked to $a$, $L_2$ is the set of elements $c$ that precede $a$ such that there is one element in the order strictly between $a$ and $c$ and so on. Given a spacetime $\mathcal{M}$ of dimension $d$ and a scalar field $\phi$ on it, for every $x\in \mathcal{M}$,  $B^{(d)}$ and the sprinkling process at density $\rho$ give rise to a random variable $\mathbf{B}_\rho\phi(x)$ which is the value of $B^{(d)}\phi(x)$ evaluated for the sprinkled causal set  and field induced on it, with an element at $x$ added by hand. 
In Minkowski space, in 2 \cite{Sorkin:unpub}
and 4 \cite{Belenchia:2015hca} dimensions, it has been proved that if $\phi$ is of compact support and if $x$ is not on the past boundary of the support of $\phi$,  the mean of this random variable, tends in the continuum limit to $\Box \phi(x)$. It should be straightforward to extend this Minkowski spacetime result to all dimensions.  In 4 dimensional curved spacetime, it has been proved that if the support of $\phi$ is a region that is small compared to any radius of curvature, the continuum limit of the mean of the 
random variable is $\Box \phi (x) - \frac{R(x)}{2}\phi(x)$ \cite{Belenchia:2015hca}.  This result 
also holds for $d=2$ but it has not been extended to other dimensions,  nor to the ``strong gravity'' or ``cosmological''  case where the size of the region is comparable to or larger than the radius of curvature. It has, however, been shown that \textit{if} the mean of the $\mathbf{B}^{(d)}\phi(x)$ random variable is a \textit{local} quantity then 
\begin{align}\label{Boxretard}
\lim_{\rho\rightarrow \infty} \langle \mathbf{B}_\rho \phi(x)\rangle = 
\Box \phi(x) - \frac{R(x)}{2}\phi(x) 
 \end{align}
 in every dimension \cite{Dowker:2013vba,Glaser:2013xha}. The coefficient $\frac{1}{2}$ of the scalar curvature term is dimension independent. 

This work on the scalar d'Alembertian, then gives a dimension dependent causal set Ricci scalar curvature analogue, by applying the operator $B$ to the constant field, -1:
\begin{equation}\label{Rretard}
\frac{1}{2} R_{causal set }^{(d)} (a) = -  \frac{1}{l^2}\left( \alpha_d  +\beta_d \sum\limits_{i=1}^{n_d}  C^{(d)}_i N_i(a) \right) \, ,
\end{equation} 
where $a$ is an element of the causal set and $N_i(a)$ is the number of elements of the causal set in 
the $i$-th level preceding $a$. As before, the sprinkling process at density $\rho$ into $\mathcal{M}$ of dimension $d$ and the causal set  function $R_{causal set }^{(d)}$ give rise to a random variable $\mathbf{R}_\rho(x)$ for 
each point $x$ of $\mathcal{M}$. 

Summing (\ref{Rretard}) over the whole causal set, multiplied by 
$\rho = l^{-d}$ for the volume element, and by the coupling $l_p^{2-d}$, then gives the causal set action (\ref{bdgaction}). 

The causal set scalar d'Alembertian and the scalar curvature analogue have \emph{advanced} versions gotten by reversing the  order in (\ref{Bretard}) and ({\ref{Rretard}) so the levels summed over are \textit{preceded by} $a$: $L_1$ is the set of elements that are {preceded by} $a$ and linked to $a$ \textit{etc.} Every result mentioned above holds, mutatis mutandis, for the advanced objects.
The final action (\ref{bdgaction}) is, however, independent of whether the advanced or the retarded version of the scalar curvature analogue is summed over the causal set to obtain it. 

The argument for our conjecture \ref{conjecture}, then goes as follows. The action is a sum over the causal set of  (\ref{Rretard}), the retarded scalar curvature estimator.
 For any point $x$ that is not strictly on the past boundary of $\mathcal{M}$, for $ \rho$  big enough, there will be enough of $\mathcal{M}$ in the past of $x$ for the value of the mean $\langle \mathbf{R}_\rho (x)\rangle$ to be $R(x)$ to as good an approximation as we like. This is because for large enough $\rho$ there is room to the past of $x$ in $\mathcal{M}$ for all the levels in the sum to fit below $x$ 
 and to get the necessary cancellations between the contributions from each level.  In particular this is the case when  $x$ lies on the future boundary and does not also lie 
on the past boundary of $\mathcal{M}$,  so we expect only the Einstein Hilbert contribution from such $x$. 
  
However, the action is invariant under order reversal. It also equals a sum over the causal set elements of the advanced version of (\ref{Rretard}). So, running the argument above for this case, when considering the mean of the random discrete action of $\mathcal{M}$ we expect only the Einstein Hilbert contribution from points $x$ that lie
on the past boundary and 
 not on the future  boundary of $\mathcal{M}$ 
 
Now, the points that are on \emph{both} the past and future boundaries, \textit{i.e.} the points of the joint, are not covered by either argument -- there 
is no spacetime to their past or their future and their contribution to the mean will not be the Einstein Hilbert contribution either way you look at it. So we expect a different contribution from the joint. On dimensional grounds, if this is a local contribution,  then it will be a dimensionless constant times the volume of the joint because, for finite $\rho$, any higher terms in the derivative expansion will appear multiplied by negative powers of $\rho$ and will go away in the limit.\footnote{In the case that the region $\mathcal{M}$ is not globally hyperbolic and has  a timelike boundary, then for similar reasons as above, we expect a contribution to the mean of the random discrete action from  the timelike boundary $\Sigma$. On dimensional grounds, if this contribution is local then in the derivative expansion the first couple of terms will be
 \begin{align}
a_1 \frac{1}{l_p^{d-2} l } \textrm{Vol}_{d-1}(\Sigma) + a_2 \frac{1}{l_p^{d-2} }\int_\Sigma
\sqrt{h} K  \,,
\end{align}
where $a_i$ are dimensionless constants and $K$ is the trace of the extrinsic curvature on 
$\Sigma$. The first term will diverge in the continuum limit and for large enough finite $\rho$ it will dominate all other terms. 
There is evidence  of this for the case of rectangles in 2d Minkowski spacetime \cite{Benincasa:2010as}. Further work on the timelike boundary is ongoing \cite{Cunningham:2020}.}

There may seem to be a contradiction between the claim that the limit of the mean of $\mathbf{R}_\rho(x)$ equals the Ricci scalar for all $x$ not on the joint 
and the claim that the limit of the mean of the action equals the Einstein Hilbert term \textit{plus} an extra boundary term. The joint is a set of measure zero after all. Where does the extra term come from? To understand this, consider  the mean of the random discrete action of a spacetime $\mathcal{M}$ at density $\rho$:
\begin{align}
\frac{1}{\hbar} <{\mathbf{S}}_\rho(M)> = 
\zeta_d \bigg[ < \mathbf{N}_\rho(M)>  + \frac{\beta_d}{\alpha_d}
 \sum_{i = 1}^{n_d} C_i ^{(d)} <\mathbf{N}_{i,\rho}(M)> \bigg]\,,
 \end{align}
where the random variables in bold, $\mathbf{N}$ and $\mathbf{N}_i$, are the cardinality of the sprinkled causal set  and the number of order intervals of cardinality $i+1$ in the sprinkled causal set,   respectively.
The means of these random variables are given by the 
Poisson distribution: 
\begin{align}\label{doubleint}
\frac{1}{\hbar} <{\mathbf{S}}_\rho(M)> = 
\zeta_d \bigg[ \rho \int\limits_{\mathcal{M}} dV   + \frac{\beta_d}{\alpha_d}
 \sum_{i = 1}^{n_d} C_i ^{(d)} \rho^2 \iint\limits_{\substack{M\times M \\ y\in J^+(x)}} dV_x\,dV_y\, \frac{(\rho V_{xy})^{i-1}}{(i-1)!} e^{- \rho V(x,y)} \bigg]\,,
 \end{align}
where $V_{xy}$ is the volume of the causal interval, $I(x,y)$, between $x$ and $y$. Consider doing the 
$y$ integration over ${\mathcal{M}}\cap J^+(x)$ first and, to emphasise the puzzle, suppose $\mathcal{M}$ is a portion of Minkowski spacetime so the result of the $y$ integral gives zero in the limit for all 
$x$ not on the boundary of $\mathcal{M}$. Taking the limit and doing the second, $x$ integration do not commute, however, 
because there are delta function-like contributions along the boundary of $\mathcal{M}$ which contribute to the  limit. We will see this explicitly in the examples analysed below.


\section{The Set Up}

Choosing an order in which to perform the  double integral (\ref{doubleint}) is 
equivalent to a choice of either the advanced or retarded version of the Ricci curvature analogue. If we do the $y$ integration first, we are choosing the advanced version:
\begin{align} 
\frac{1}{\hbar} <{\mathbf{S}}_\rho(M)> &= 
\frac{1}{l_p^{d-2}}   \int_{\mathcal{M}} dV_x \mathcal{L}_\rho (x)\,, \label{meanS}\,\\ 
{\textrm{where}}\quad\quad  \mathcal{L}_\rho(x) &:=  -\rho^{\frac{2}{d}}\Big(\alpha_d   + \rho\, \beta_d  \int\limits_{\substack{J^+(x)\cap \mathcal{M} }} dV_y \, {\mathcal{O}}_d \,e^{-\rho V_{xy}} 
\Big)\,, \\
{\textrm{and}}\quad\quad  {\mathcal{O}}_d& := \sum_{i=1}^{n_d} \frac{C_i^{(d)}}{(i-1)!} \rho^{i-1}
 (-\frac{d}{d\rho})^{i-1} \,.
 \end{align}
Introducing the differential operator ${\mathcal{O}}_d$ makes the formulae simpler to write and also 
goes some way to explaining why such expressions have a hope of giving finite answers since one can show that 
${\mathcal{O}}_d$ annihilates certain powers of $\rho$ that would otherwise make the expression divergent in the limit. 

We will test Conjecture \ref{conjecture} by calculating (\ref{meanS})  in the limit of $\rho \rightarrow \infty$ for some examples of regions in a conformally flat spacetime in 2 and 4 dimensions, to first order in a curvature expansion.

\subsection{Metric}

The metric in all examples we will consider is conformally flat,
\begin{align} \label{metric}
ds^2  = \Phi^2(t) \eta_{\mu\nu} dx^\mu dx^\nu =  \Phi^2(t) (- dt^2 + \delta_{ij} dx^i dx^j)\,,
\end{align}
with a simple conformal factor $\Phi(t) = 1 + b t^2$ where $b$ is a constant. 
At $t=0$, in $d$ dimensions the Ricci curvature components are 
\begin{align}
R_{00} &= - 2(d-1) b\,,  \\
R_{ii} &=  2 b\,, \\
R_{0i} & = 0 \,, \\
R &= 4(d-1) b \,.
\end{align}
The height in $t$ of the regions we will consider will be
small compared to $b^{-\frac{1}{2}}$ and the curvature components will 
be approximated as constant throughout the region. In the calculations below we will assume that there are no divergences arising from the higher order curvature terms which are simply dropped in the calculation whenever they arise.  
We will not bother to write ``$+ \dots$'' to indicate that higher order curvature terms have been dropped. 

We will need the proper time, $\tau_{xy}$ between two points $x$ and $y$ in the spacetime and the 
volume, $V_{xy}$, of the causal interval, $\mathcal{I}(x,y)$, between them  ($x \in J^-(y)$). 

\subsection{$\tau_{xy}$}

Let $X^\mu$ and $Y^\mu$ be the coordinates of points $x$ and $y$ respectively. 
Let $\Delta^\mu = Y^\mu - X^\mu$. 
The geodesic $x^\mu(\tau)$ from $x$ to $y$ satisfies
\begin{align}
\frac{dx^i}{d\tau} &= c^i (1 + bt^2)^{-2} \\
\frac{dt}{d\tau} &= (1 + bt^2)^{-2}  [c^2 + (1+bt^2)^2]^\frac{1}{2}
\end{align} 
where $c^i$ is a constant and $c^2 = ||c^i||^2$ and 

\begin{align}
c^i &= \gamma \Delta^i  \\
\gamma &= \tau_{0xy}^{-1} ( 1 + b  \frac{1}{3} \frac{((Y^0)^3 - (X^0)^3 )}{\Delta^0})  \\
\tau_{0xy}^2& = (\Delta^0)^2 - ||\Delta^i ||^2 \,.
\end{align} 
$\gamma$ is given to first order in curvature. 
From this we 
find
\begin{align}
\tau_{xy}& = \tau_{0xy} \left( 1 +   \frac{b}{3} \frac{((Y^0)^3 - (X^0)^3 )}{\Delta^0}\right)\\
&= \tau_{0xy} \left( 1 +  \frac{b}{3}((Y^0)^2 + Y^0 X^0 +  (X^0)^2 )\right)\,.
\end{align}

\subsection{$V_{xy}$}

We can either calculate $V_{xy}$ directly or use the formula (74) from \cite{Gibbons:2007nm}. 
Although the formula is expressed in Riemann normal coordinates (RNC), the proper time $\tau_{xy}$ is a coordinate invariant and the same formula holds in conformally flat coordinates to first order in curvature. We have
\begin{align}
V_{xy} = \Omega_{d-2} \frac{ 2^{1-d} }{d(d-1)} \tau_{0xy}^d
\Bigg(1 & + \frac{bd}{3} ( (Y^0)^2 + Y^0 X^0 +  (X^0)^2 )\\
& -\frac{bd}{24(d+1)(d+2)}  ( 6d \tau_{0xy}^2 + 2(d^2-4)(Y^0- X^0)^2 ) \Bigg)\,,
\end{align}
where $\Omega_{d-2}$ is the volume of the $(d-2)$-sphere.

\section{2 dimensions}

We will look at three examples in 2 dimensions:  a causal diamond, a slab of a cylinder, and a null triangle. For $d=2$ we have 
$R = 4 b$
and 
\begin{align}
V_{xy} = \frac{1}{2} \tau_{0xy}^2
\Big(1 & + \frac{b}{12}  ( - \tau_{0xy}^2 + 8(Y^0)^2 + 8(X^0)^2 + 8X^0Y^0 ) \Big) \,,
\end{align}
and 
\begin{align} 
\frac{1}{\hbar} <{\mathbf{S}}_\rho(M)> &= 
   \int_{\mathcal{M}} dV_x \mathcal{L}_\rho(x)\,, \label{meanS2d}\,\\ 
{\textrm{where}}\quad\quad  \mathcal{L}_\rho(x) &:=  2 \rho \Big( 1  - 2 \rho
  \int\limits_{\substack{J^+(x)\cap \mathcal{M} }} dV_y  \, {\mathcal{O}}_2 \, e^{-\rho V_{xy}} 
\Big)\,,\label{eq:yintegral} 
 \end{align}
and $ {\mathcal{O}}_2 :=  
1 + 2\rho \frac{d}{d\rho}  + \frac{1}{2}\rho^2 \frac{d^2}{d\rho^2} $.

 To perform the first $y$ integral,  the exponential in the integrand can be expanded in curvature and terms quadratic and higher in $b$ dropped:
\begin{align}
V_{xy} &= V_{0xy} + \delta V_{xy}\,,\\
V_{0xy} &= \frac{1}{2} \tau_{0xy}^2\,,\\
 \delta V_{xy}& = \frac{b}{24} \tau_{0xy}^2 (  - \tau_{0xy}^2 + 
8(Y^0)^2 + 8(X^0)^2  + 8X^0Y^0) \,,
\end{align}
and 
\begin{align}
e^{-\rho V_{xy}}  &= e^{-\rho V_{0xy} } e^{-\rho  \delta V_{xy}}\\
&=  e^{-\rho V_{0xy} }( 1 -\rho  \delta V_{xy}) \,.
\end{align}

The $y$ integral in (\ref{eq:yintegral}) is 
\begin{align}
\int\limits_{J^+(x)\cap {\mathcal{M}}}
 d^2y \, {\mathcal{O}}_2 \Bigg[ \Big(1 + 2b (Y^0)^2 - \rho  \frac{b}{24} \tau_{0xy}^2 (  - \tau_{0xy}^2 + 
8(Y^0)^2 + 8(X^0)^2  + 8X^0Y^0) \Big) e^{-\frac{\rho}{2} \tau_{0xy}^2} \Bigg]\,.
\end{align}

In all three cases we will calculate $L_\rho(x)$ and check that it tends to $\frac{R}{2}=2b$ in the continuum limit for every $x$ that is not on the future boundary of $\mathcal{M}$. There 
will also be terms in $L_\rho(x)$ that behave like delta functions on the future boundary of $\mathcal{M}$ in the limit. Those terms 
 must be integrated over $\mathcal{M}$ and the limit taken to see what contribution if any they give to the mean of the action. As described above in the section on  general $d$, doing the double integral in the other order, using the retarded form of the integrand instead of the advanced, 
 would give an $L_\rho(y)$ that is an Einstein Hilbert term plus terms with distributional behaviour on the 
 \textit{past} boundary. This leads to the expectation that the only boundary contribution actually comes from the joint.

\subsection{The interval}

Let  $\mathcal{M}$ be the causal interval, $I{(p,q)}$, centred at the origin with endpoints 
$p$ at $(-\frac{T}{2}, 0)$ and $q$ at $(\frac{T}{2},0)$ -- shown in Figure \ref{fig:2ddiamond} --  with metric (\ref{metric}) for $d=2$. 
\begin{figure}[h!]
\centering
{\includegraphics[scale=0.27]{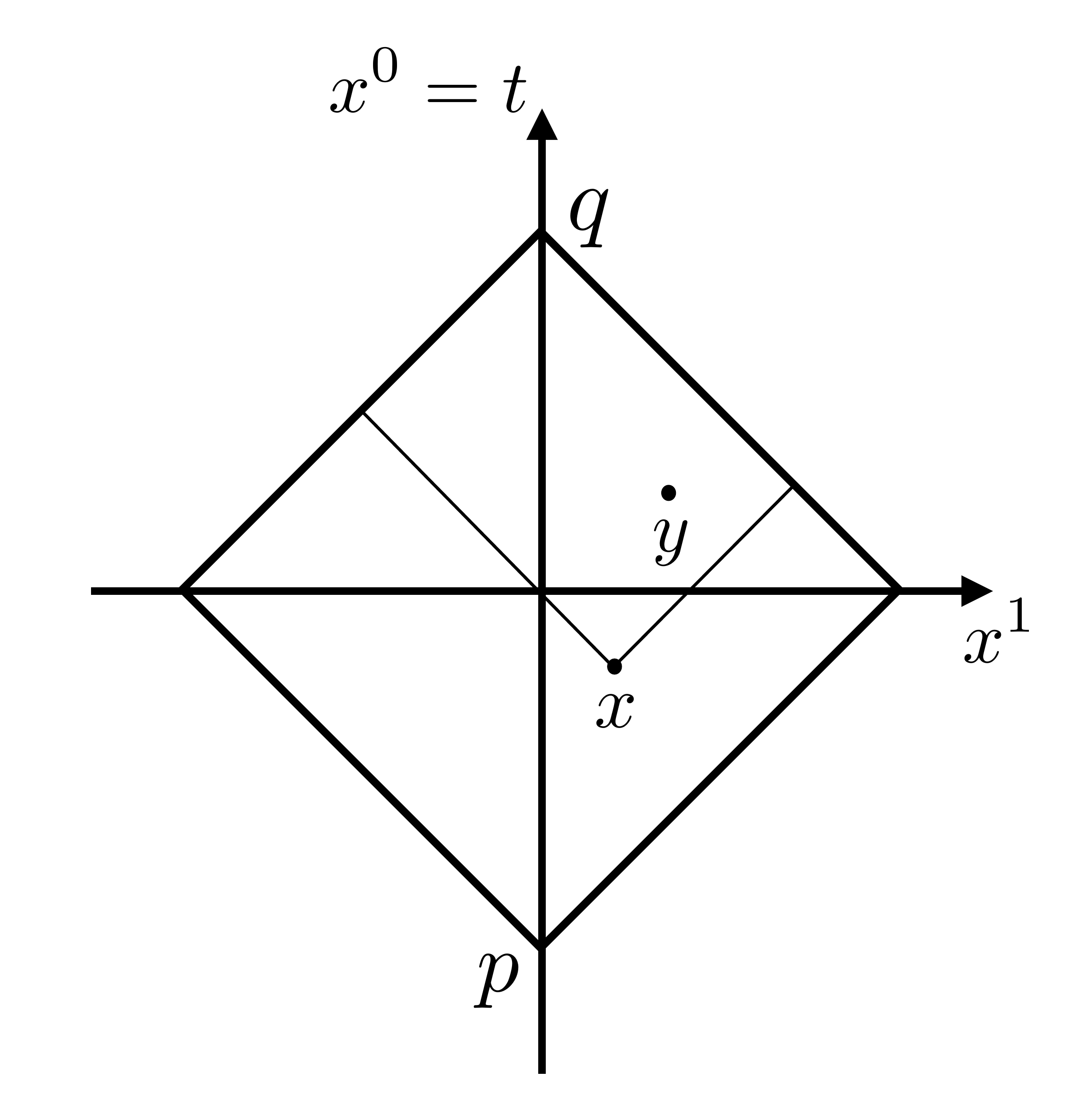}}
\caption{\label{fig:2ddiamond} 
The causal interval $I{(p,q)}$ between $p$ at $(-\frac{T}{2}, 0)$ and $q$ at $(\frac{T}{2},0)$. 
}
\end{figure}
The interval has an $S^0$ joint with volume equal to 2.

The boundaries of the regions of integration are straightforward because null geodesics in our conformally flat coordinates are straight lines as in Minkowski space. For the $y$ integral, it is  convenient to change coordinates to null coordinates, $u$ and $v$ centred at point $x$ in terms of which we have $Y^0 = X^0 + \frac{(u+v)}{\sqrt{2}}$ and $\tau_{0xy}^2 = 2 u v$. 
We find
\begin{align}
L_\rho(x) &= 2 b-2 b e^{-\frac{\rho \tau^2}{2}}-b e^{-\frac{\rho \tau^2}{2}}
\rho \tau^2 -e^{-\frac{\rho \tau^2}{2}} \rho^2 \tau^2+2 e^{-\frac{\rho \tau^2}{2}} \rho\\
&{-\frac{1}{12} b e^{-\frac{\rho \tau^2}{2}} \rho^2 \tau^4-\frac{1}{24} b e^{-\frac{\rho \tau^2}{2}} \rho^3 \tau^6}\\
&{-4 b e^{-\frac{\rho \tau^2}{2}} \rho^2
\tau^2 {X^0}^2+b e^{-\frac{\rho \tau^2}{2}} \rho^3 \tau^4 {X^0}^2}
\\
&{+\frac{1}{3} b e^{-\frac{\rho \tau^2}{2}} \rho^3
{(\frac{T}{2} - X^0)}^2 \tau^4-\frac{4}{3} b e^{-\frac{\rho \tau^2}{2}} \rho^2 {(\frac{T}{2} - X^0)}^2
\tau^2}\\
&{-4 b e^{-\frac{\rho \tau^2}{2}} \rho^2 {(\frac{T}{2} - X^0)}
\tau^2 {X^0}+b e^{-\frac{\rho \tau^2}{2}} \rho^3 {(\frac{T}{2} - X^0)} \tau^4 {X^0}}\,,
\end{align} 
where $\tau^2 = (\frac{T}{2} - X^0)^2 - (X^1)^2$ is the square of the Minkowski proper time from 
$x$ to $q$. $\tau = 0$ if and only if $x$ lies on the future boundary of $\mathcal{M}$. We see that the continuum limit of $L_\rho(x)$ is $2b$ for all $x$ not on the future boundary of $\mathcal{M}$. The first term $2b$ in $L_\rho(x)$ will give the Einstein Hilbert action when integrated over $\mathcal{M}$. The other terms have factors that tend to delta functions or derivatives of delta functions on the future boundary of $\mathcal{M}$ in the $\rho\rightarrow \infty$ limit. 

The $x$ integral can be done using null coordinates, $(u,v)$ centred, now, at $q$ in 
which $\tau^2 = 2 uv$ and the result is
\begin{align}
\frac{1}{\hbar} <{\mathbf{S}}_\rho(M)> = \frac{1}{12 \rho}&\Big(96 b-96 b e^{-\frac{\rho T^2}{2}}-96 b \gamma  +24 \rho-24 e^{-\frac{\rho T^2}{2}}
\rho+12 b \rho T^2\\
& -12 b e^{-\frac{\rho T^2}{2}} \rho T^2+b e^{-\frac{\rho T^2}{2}} \rho^2 T^4-96 b \Gamma[0,\frac{\rho
T^2}{2}]\\
&-96 b \log[\rho T^2/2]\Big)\,.
\end{align}
The limit is 
\begin{align}
\lim_{\rho\rightarrow \infty} \frac{1}{\hbar} <{\mathbf{S}}_\rho(M)> = bT^2 + 2\,.
\end{align}
The Einstein Hilbert action equals  $\frac{R}{2}\times \frac{T^2}{2} = bT^2$, 
and the volume of the joint equals 2, so this agrees with the conjecture. 

\subsection{The slab}

Now let $\mathcal{M}$ be the spacetime with metric (\ref{metric}) for $d=2$ for $0 \le t \le T$ and with $(t, x^1)$ identified with $(t, x^1 + L)$ so space is a circle. $T < 2L$ so there is no wrap-around of causal intervals in $\mathcal{M}$ and $bT^2 \ll 1$. 
\begin{figure}[h!]
\centering
{\includegraphics[scale=0.27]{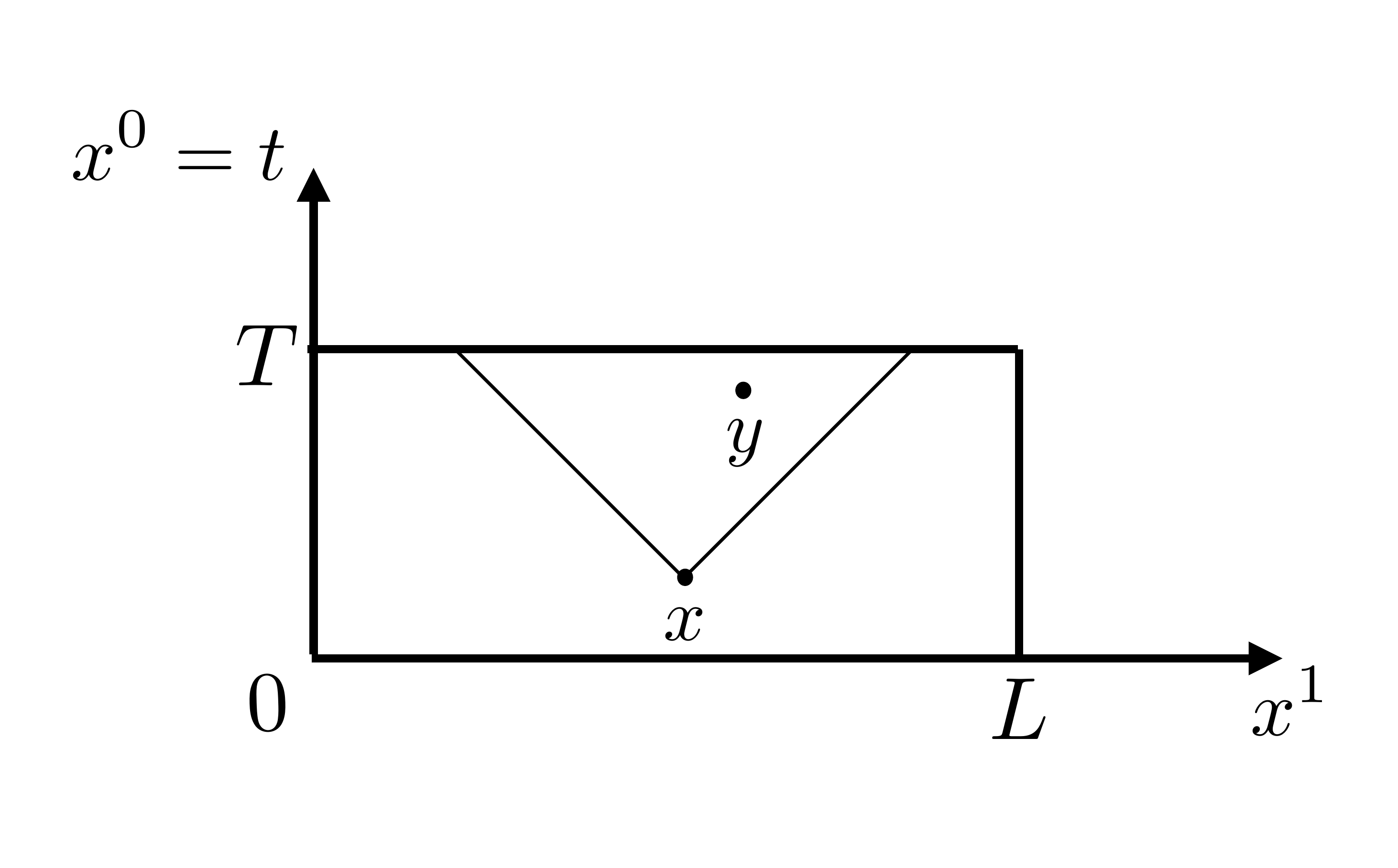}}
\caption{\label{fig:2dslab} 
A slab of a cylinder spacetime. The first, $y$ integral is over the region in the slab in the causal future of 
$x$. 
}
\end{figure}
There is a spacelike past boundary and a spacelike future boundary but  there is 
no joint. 

We find
\begin{align}\label{eq:slabell}
L_\rho(x) 
 = 2 \rho+\frac{1}{24} Q \rho \Big[&-24 Q \rho+24 b {X^0}+b Q \big(-3+\rho (7 Q^2 (-8+Q^2 \rho)\\
 &+24
Q (-6+Q^2 \rho) {X^0}+24 (-5+Q^2 \rho) {X^0}^2)\big)\Big]\\
-\frac{1}{12 \sqrt{2}}\sqrt{\rho}&\textrm{DawsonF}[\frac{Q \sqrt{\rho}}{\sqrt{2}}]
\Big[-24 Q \rho (-3 +Q^2 \rho)+24 b {X^0}\\
&+b Q (-3+\rho (Q^2 (39+7 Q^2 \rho (-9+Q^2 \rho))\\
&\quad\quad+24
Q (5+Q^2 \rho (-7+Q^2 \rho)) {X^0}\\
&\quad\quad
+24 (3+Q^2 \rho (-6+Q^2 \rho)) {X^0}^2))\Big]\,,
\end{align}
where $Q := T - X^0$. One cannot just read off the limit as one could for the  interval, but $L_\rho(x)$ does have the correct limit of $2b$ as $\rho\rightarrow\infty$ for every $x$ for
which $Q \ne 0$ \textit{i.e.} for every $x$ not on the future boundary of $\mathcal{M}$. In the expression for  $L_\rho(x)$ there are individual terms that have a distributional character at $Q=0$ in the limit, but  integrating $L_\rho(x)$ over $\mathcal{M}$ and taking the limit we find that these all cancel and we get 
\begin{align}
\lim_{\rho\rightarrow \infty} \frac{1}{\hbar} <{\mathbf{S}}_\rho(M)> = 2bLT\,.
\end{align}
The Einstein Hilbert action equals $\frac{1}{2} R \times LT = 2 b L T$ to first order in curvature and 
there is no joint so this agrees  
with the conjecture.

\subsection{The triangle} 

Now let $\mathcal{M}$ be the null triangle, or half-interval,  shown in Figure \ref{fig:2dtriangle},
 with apex $p$ at the origin, ``base'' at $x^0 = T$, and two null boundary segments. 
\begin{figure}[h!]
\centering
{\includegraphics[scale=0.27]{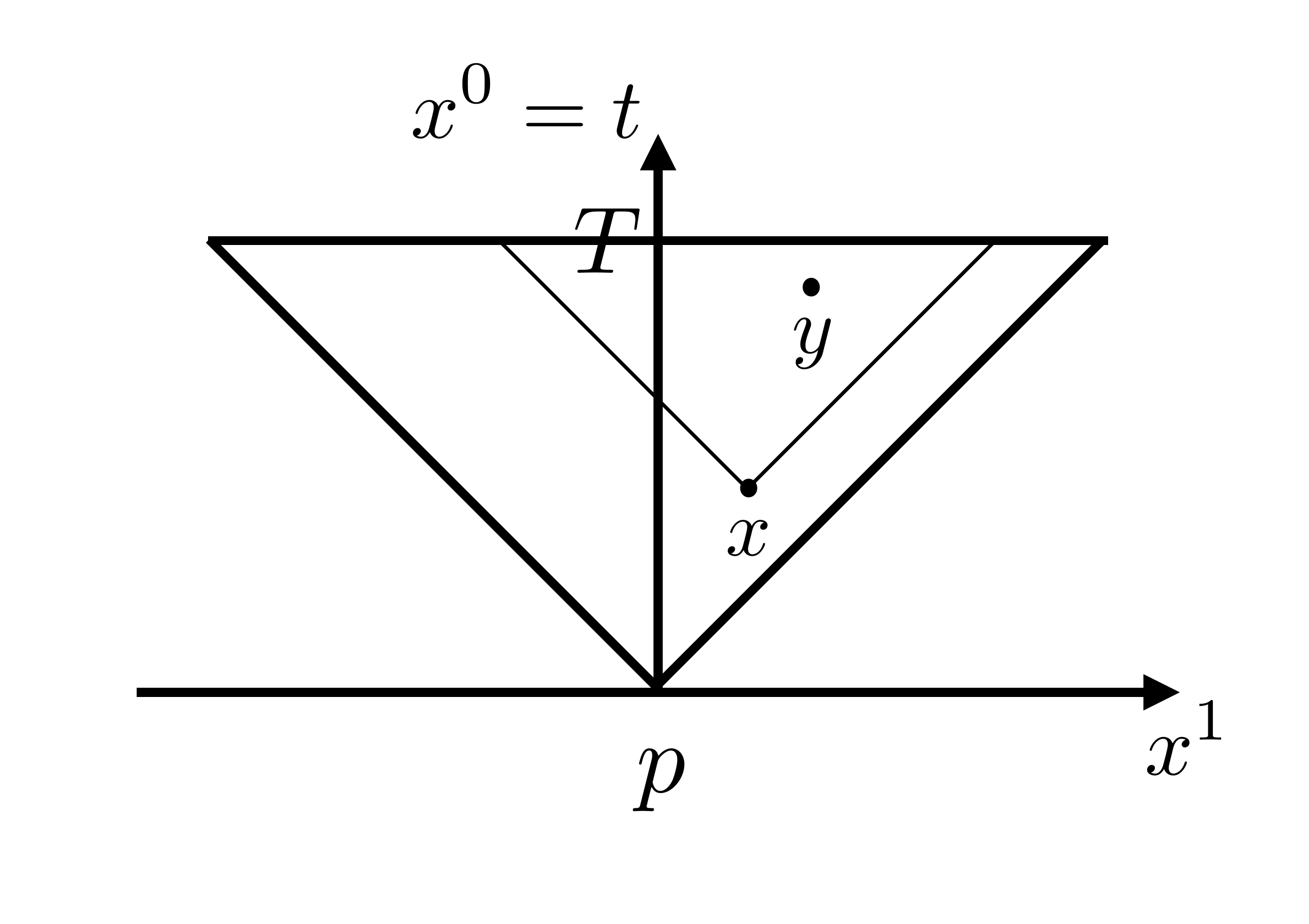}}
\caption{\label{fig:2dtriangle} 
The null triangle of coordinate height $T$. 
}
\end{figure}
The joint is an $S^0$.

The $y$ integral is the same as for the slab, and $L_\rho(x)$ equals (\ref{eq:slabell}). Integrating this over the triangle and taking the limit gives,
\begin{align}
\lim_{\rho\rightarrow \infty} \frac{1}{\hbar} <{\mathbf{S}}_\rho(M)> = 2+ 2bT^2\,.
\end{align}
The Einstein Hilbert action equals $\frac{R}{2} \times T^2 = 2b T^2$ and the volume of the joint equals 2, so this agrees with the conjecture. 

\section{ 4 dimensional causal interval}

We take $\mathcal{M}$, to be the causal interval,  the 4-dimensional  analogue of Figure \ref{fig:2ddiamond}, upright, centred on the origin and of coordinate height $T$, with past and future endpoints $p$ and $q$ respectively. The joint lies in the $t=0$ plane by symmetry and the volume of the joint is the volume of a $2$-sphere of radius $T/2$,
$
Vol_2(J) =  \pi T^2\,.
$

In 4 dimensions we have $R = 12b$ and 
\begin{align}
V_{xy}&  = \frac{\Omega_{2} }{8.4.3} \tau_{0xy}^4
\left(1  + \frac{4b}{3}  \frac{((Y^0)^3 - (X^0)^3 )}{\Delta^0}
  -\frac{2b}{15}  ( \tau_{0xy}^2 + (Y^0- X^0)^2 ) \right)\\
& = \frac{\pi}{24} \tau_{0xy}^4\left(1  + \frac{20b}{15} ((X^0)^2 + (Y^0)^2 + X^0Y^0)
  -\frac{2b}{15}  ( \tau_{0xy}^2 + (Y^0- X^0)^2 )  \right)\\
& = \frac{\pi}{24} \tau_{0xy}^4\left(1  + \frac{2b}{15} (  - \tau_{0xy}^2 + 
9(Y^0)^2 + 9(X^0)^2  + 12X^0Y^0) \right)
\end{align}
and
\begin{align}
\frac{1}{\hbar} <S> &= \frac{1}{l_p^2} \int_{\mathcal{M}} d^4x \sqrt{-g(x)} L_\rho (x)\,, 
\\
L_\rho (x)  &= \frac{4}{\sqrt{6}}  \rho^{\frac{1}{2}} \Big(1   -  \rho  {\mathcal{O}}_4
 \int_{J^+(x)\cap {\mathcal{M}}} d^4 y \sqrt{-g(y)} e^{-\rho V_{xy}} \Big)\,, 
  \\
{\mathcal{O}}_4& = 1 + 9\rho \frac{d}{d\rho} + 8 \rho^2 \frac{d^2}{d\rho^2} + \frac{4}{3} \rho^3 \frac{d^3}{d\rho^3} \,.
\end{align}

 To perform the first $y$ integral,  the exponential in the integrand is expanded in curvature and terms quadratic and higher in $b$ dropped, as before:
\begin{align}
\int\limits_{J^+(x)\cap {\mathcal{M}}}
 d^4y \, {\mathcal{O}}_4 \left[ \big(1 + 4b (Y^0)^2 - b  \frac{\rho\pi}{180} \tau_{0xy}^4 (  - \tau_{0xy}^2 + 
9(Y^0)^2 + 9(X^0)^2  + 12X^0Y^0) \big) e^{-\frac{\rho\pi}{24} \tau_{0xy}^4} \right]\,.
\end{align}

The $y$ integral is over the causal interval between $x$ and $q$ and details of the calculation are given in Appendix \ref{appendix}. The result for $L_\rho(x)$ is 
\begin{align} 
L_\rho(x) = &+6 b \text{Erf}[\frac{1}{2} \sqrt{\frac{\pi }{6}}
\sqrt{\rho} {\tau}^2] \label{eq:one}\\
&+ 2 \sqrt{\frac{2}{3}} e^{-\frac{1}{24} \pi  \rho {\tau}^4} \sqrt{\rho}
 +\frac{4}{5} \sqrt{\frac{2}{3}} b e^{-\frac{1}{24}
\pi  \rho {\tau}^4} \sqrt{\rho} {T^0}^2 +\frac{4}{5} \sqrt{\frac{2}{3}}
b e^{-\frac{1}{24} \pi  \rho {\tau}^4} \sqrt{\rho} {\tau}^2
\label{eq:two}\\
&-\frac{b e^{-\frac{1}{24} \pi  \rho {\tau}^4} \pi  \rho^{3/2}
{T^0}^2 {\tau}^4}{5 \sqrt{6}} +\frac{b e^{-\frac{1}{24} \pi  \rho {\tau}^4} \pi  \rho^{3/2} {\tau}^6}{45 \sqrt{6}}
\label{eq:three}\\
&-\frac{1}{3}
\sqrt{\frac{2}{3}} b e^{-\frac{1}{24} \pi  \rho {\tau}^4} \pi  \rho^{3/2} {T^0} {\tau}^4 (\frac{T}{2} -T^0)-\frac{1}{3} \sqrt{\frac{2}{3}}
b e^{-\frac{1}{24} \pi  \rho {\tau}^4} \pi  \rho^{3/2} {\tau}^4 (\frac{T}{2} -T^0)^2 
\label{eq:four}\\
& +\frac{128 \sqrt{6} b {T^0}^2}{5 \pi  \sqrt{\rho} {\tau}^4} 
+\frac{112 \sqrt{6}
b e^{-\frac{1}{24} \pi  \rho {\tau}^4} {T^0}^2}{5 \pi  \sqrt{\rho} {\tau}^4}-\frac{288 b {T^0}^2 \text{Erf}[\frac{1}{2} \sqrt{\frac{\pi }{6}} \sqrt{\rho} {\tau}^2]}{\pi
 \rho {\tau}^6}
\label{eq:five}\\
& -\frac{192 \sqrt{6} b}{5 \pi  \sqrt{\rho}
{\tau}^2} +\frac{132 \sqrt{6} b e^{-\frac{1}{24} \pi  \rho {\tau}^4}}{5 \pi  \sqrt{\rho} {\tau}^2}+\frac{72 b \text{Erf}[\frac{1}{2} \sqrt{\frac{\pi }{6}} \sqrt{\rho} {\tau}^2]}{\pi  \rho {\tau}^4}\,,\label{eq:six}
\end{align}
where $\tau^2 := (\frac{T}{2} - X^0)^2 - ||X^i||^2$ is the square of the Minkowski proper time from 
$x$ to $q$, and $T^0:= \frac{T}{2} - X^0$. 
As $\rho\rightarrow \infty$, for non-zero $\tau$, the first term tends to $6b$, the Einstein Hilbert term,
whilst all other 
terms tend to zero. $\tau=0$ is the future boundary of $\mathcal{M}$ and on that boundary, many of the terms of $L_\rho(x)$ have a distributional behaviour in the limit.  Integrating $L_\rho(x)$ over $\mathcal{M}$ -- see Appendix \ref{appendix} --  and taking the limit we find
\begin{align}
\lim_{\rho\rightarrow \infty} \frac{1}{\hbar} <{\mathbf{S}}_\rho(M)> = \frac{1}{l_p^2}(6b\times \textrm{Vol}(M) + \pi T^2)\,.
\end{align}
$ 6b\times \textrm{Vol}(M)$ is the Einstein Hilbert action and $\pi T^2$ is the area of the joint, 
in agreement with the conjecture.

\section{Discussion}

There is much further work to be done on the conjecture. 
It is not easy to see from the calculations above that the boundary contribution is concentrated at the joint. For the $d=4$ interval, the integrand of the second integral $L_\rho(x)$ has many terms that are individually distributional in the limit and the joint volume does not come from any single one of them. It may be possible to analyse the integrand of the full double integral over $x$ and $y$ and, without choosing the advanced or retarded order of integration, see that there is a distributional character to that integrand that is an appropriate delta function on the joint only. This could help in proving the conjecture in more generality. 
In working to first order in curvature we are essentially assuming that the limit is local and has a 
derivative expansion. To be rigorous, the higher order terms should be bounded and shown to tend to zero 
in the limit.  

One stumbling block is our lack of knowledge about the causal set scalar d'Alembertian away from the low curvature regime. If we knew that the result (\ref{Boxretard}) held in general then it would greatly strengthen the conjecture at least as far as the Einstein Hilbert term is concerned. 
As well as analytic work on more examples, one could do simulations to gain evidence one way or 
another. There exist generalisations of the causal set actions that are more non-local and arise from a
generalisation of the causal set scalar d'Alembertian that employs an averaging over many layers akin to a smeared ``blocking'' on a lattice \cite{Sorkin:2007qi}. This smearing helps dampen the large fluctuations in the causal set action and is therefore useful for simulations. Even with this, however, the large size of causal sets required makes it difficult to ascertain when the asymptotic regime has been reached \cite{Benincasa:thesis}.

 If the conjecture holds, the causal set action of a manifold-like causal set is -- up to fluctuations which we are ignoring  -- approximately local both in its bulk and boundary terms.  Whereas, the action of a non-manifold-like causal set is nonlocal and, because there is no cancellation between the numbers of order intervals, is typically of order of $N^2$ where $N$ is the cardinality of the causal set. 
This paints a heuristic picture of how a path integral over causal sets could conquer the entropic weight of the vastly  more numerous non-manifold-like causal sets and pick out the ones that 
have continuum approximations.  In the path sum, we require (i)  there is a continuum regime \textit{i.e.} non-manifold-like causal sets are suppressed and (ii) non-GR solutions are suppressed. 
For the second requirement, assuming there is a continuum regime,  we want the action to pick out the solutions of the Einstein equations and the causal set action being close to the Einstein-Hilbert action
for a manifold-like causal set is a promising sign, though it may be necessary also to add extra boundary terms to the action \cite{Buck:2015oaa}.  The first requirement is tantamount to solving one aspect of the cosmological constant problem: why is there a continuum regime at all if quantum gravity has no free parameters and only one fundamental scale? Stationary phase heuristics suggest that a causal set will 
be suppressed in the sum if small changes in the causal set 
cause large changes in the action. When the causal set is non-manifold-like, the action is huge and 
so a small change in the causal set will indeed cause a large variation in the action. Also, in the continuum regime, the dominance of the timelike boundary term in the random discrete  action might act to suppress causal sets with timelike boundaries and similarly the -- albeit much slower, logarithmic -- divergent behaviour of the random discrete action for the trousers might act to suppress such topology changes. 
Whether or not these heuristics are a good guide in a discrete theory like causal set theory,  
it seems unlikely that the theory of quantum causal sets will be based fundamentally on the BDG action 
because of its dimension dependence: quantum gravity should explain $d=4$, not put it in by hand. 
Quantum causal sets will more likely be governed fundamentally by something like a quantal version of the classical sequential growth models \cite{Rideout:2000a,Sorkin:2011sp,Dowker:2010qh,Surya:2020cfm}.  
Nevertheless, one can imagine the causal set action being relevant in some intermediate regime of the
theory --  between the fundamental and the continuum regimes --  and it is important to study how path sums defined using the action behave. 
Such causal set path sums are beginning to be investigated in $d=2$ and $d=3$ \cite{Surya:2011du,Glaser:2017sbe,Cunningham:2019rob}. 

 If Conjecture \ref{conjecture} holds, it  would give the value of the continuum limit of the mean of the  ``spacetime mutual information'' (SMI) in the case when the spacetime to the past of a Cauchy surface, $\Sigma$, is divided into two by a horizon, $H$ \cite{ Benincasa:thesis, Dowker:kitp}. 
 The SMI in this case equals the  sum of the actions of the  interior of the horizon and the exterior of the horizon minus the action of their union (the whole spacetime). The SMI is nonzero due to the bilocal nature of the action and, if Conjecture \ref{conjecture} holds,  then the Einstein Hilbert terms cancel and the  limiting value of the mean of the SMI is equal to the area of the intersection of the Cauchy surface and the horizon, in fundamental units: $\frac{\textrm{Vol}_{d-2}(\Sigma\cap H)}{l^{d-2}}$.
 
 Another, related, consequence of Conjecture \ref{conjecture} is that if the manifold $\mathcal{M}$ has no joint and is  divided into $\mathcal{M}^-$ and $\mathcal{M}^+$, the past and future respectively of 
a Cauchy surface, $\Sigma$, that does not intersect the past or future boundary of $\mathcal{M}$,  then the limit of the mean of the discrete random action is additive because it is the Einstein Hilbert action, for each of $\mathcal{M}$, $\mathcal{M}^-$ and $\mathcal{M}^+$. In ordinary quantum mechanics, additivity of the action translates into the so-called folding property of the path integral propagator. At finite $\rho$ there are contributions to the mean action of $\mathcal{M}$ that are bilocal and straddle $\Sigma$, but these become weaker as the sprinkling density becomes larger. The relevance of these observations to 
the causal set path sum remain to be explored. 
  
 If the conjecture turns out to fail, we can  hope it  fails in an interesting and comprehensible way. 

\section{Acknowledgments} I thank Ludovico Machet and Jinzhao Wang for sharing with me their calculations of the mean of the discrete random action of Riemann normal neighbourhoods which stimulated this work. The results on causal intervals are a special case of their general result on Riemann normal neighbourhoods in all dimensions \cite{1808935}. We have used different methods both of which can be useful in future work and have agreed to publish both sets of calculations. I also thank Sumati Surya and Ian Jubb for useful comments. This
research was supported in part by Perimeter Institute for Theoretical Physics. Research at Perimeter Institute is supported by the Government of Canada through
Industry Canada and by the Province of Ontario through the Ministry of Economic Development and Innovation. FD is supported in part by STFC grant
ST/P000762/1 and Royal Society APEX grant APX/R1/180098. 

\appendix
\section{Calculations for $d=4$}\label{appendix}

To do the $y$ integration over $I(x,q)$, it is convenient to change to null radial coordinates centred at $x$ and in which $I(x,q)$ is an 
upright interval.  We do this via a series of coordinate transformations, 
all Poincare transformations.
First translate the origin to $x$. Then rotate in space so that the only non-zero spatial coordinate
of $q$ is in the positive 1-direction. Finally, boost in the $1$-direction so 
that $I(x,q)$ is upright. By properties of translations and boosts, the ``coordinate
proper time" between the two endpoints of the interval does not change, it always equals $\tau_{0xq}$ where
\begin{align}
\tau_{0xq}^2 = (\frac{T}{2} - X^0)^2 - ||X^i||^2\,.
\end{align}
Here and elsewhere we use the notation $||X^i||$ for the Euclidean norm of the vector with components $X^i$.
For the purposes of the $y$ integral, $\tau_{0xq}$ is a constant because $X^\mu = (X^0, X^i)$ are constants. 

Call the new coordinates in which $x$ is at the origin and $I(x,q)$ is upright  $\{z^\mu\}$. The metric is still conformally flat in these coordinates 
and null geodesics remain straight lines at 45 degrees in the new coordinates. 

The coordinate height of the upright interval is $\tau_{0xq}$. The point $x$ is at the origin of the 
$z^\mu$ coordinates and $q$ is at $(\tau_{0xq}, 0, 0, 0)$. 

We need $Y^0$ and $\tau_{0xy}$ in the 
new coordinates:
\begin{align}
Y^0 &= \gamma (z^0 + w z^1) + X^0\,,\\
\tau_{0xy}^2 &= (z^0)^2 - ||z^i||^2 \,,
\end{align}
where 
\begin{align}
w &= r_X (\frac{T}{2} - X^0)^{-1} \,,\\
r_X & = || X^i ||\,,\\
\gamma & = (1 - w^2)^{-\frac{1}{2}}\,.
\end{align}

Finally we define null radial coordinates with origin at point $x$: 
\begin{align}
u& = \frac{1}{\sqrt{2}}(z^0 - ||z^i||)\,,\\
v &= \frac{1}{\sqrt{2}}(z^0  + ||z^i||)\,,
\end{align}
together with polar angles $\theta$ and $\phi$. 
Let us also choose the the polar angles so that $ z^1 = ||z^i|| \cos \theta $. 

 $I(x,q)$ is given by the ranges
\begin{align}
v &\in [0,\frac{1}{\sqrt{2}} \tau_{0xq} ]\,,\\
u& \in  [ 0,v] 
\end{align}
and
\begin{align}
d^4y = dv\, du \, \frac{1}{2} (v-u)^2 d\Omega_2 \,.
\end{align}

The $y$ integral is then

\begin{align}
&\int_{I(x,q)} d^4 y \, {\mathcal{O}}_4  \Big[e^{-\rho V_{0xy}}  \Big(1 + 4 b (Y^0)^2
 \\
& \quad \quad\quad  - \rho \,b\, \frac{\pi}{24} \tau_{0xy}^4  \frac{2}{15} (  - \tau_{0xy}^2 + 
9(Y^0)^2 + 9(X^0)^2  + 12X^0Y^0)\Big)\Big]
\\
& = \int_0^{\tau_{0xq}/\sqrt{2}} dv \int_0^v du  \int_0^\pi d\theta\sin\theta\int_0^{2\pi} d\phi \frac{(v-u)^2}{2}    {\mathcal{O}}_4 \Big[e^{- \rho \frac{\pi}{6}u^2v^2 }
\\
&\quad\quad \Big( 1 +  4b(Y^0)^2  - \rho\,b\, \frac{\pi}{45}( uv)^2 [ -2uv + 9(X^0)^2 + 9(Y^0)^2 + 12X^0 Y^0] \Big)\Big]\,,
\end{align}
with 

\begin{align}
Y^0 &= \gamma (z^0 + w z^1) + X^0\,,
\end{align}
where 
\begin{align}
z^0&= \frac{(u+v)}{\sqrt{2}}\,,\\
z^1& =  \frac{(u-v)}{\sqrt{2}}\cos\theta\,,\\
w &= {T^0}^{-1}  ((T^0)^2 - \tau_{0xq}^2)^{\frac{1}{2}} \,,\\
\gamma & = T^0\tau_{0xq}^{-1}\,,\\
T^0 &= \frac{T}{2} - X^0\,.
\end{align}
Mathematica can perform the $y$ integral and this gives 
\begin{align} 
L_\rho(x) = &+6 b \text{Erf}[\frac{1}{2} \sqrt{\frac{\pi }{6}}
\sqrt{\rho} {\tau}^2] \label{eq:one}\\
&+ 2 \sqrt{\frac{2}{3}} e^{-\frac{1}{24} \pi  \rho {\tau}^4} \sqrt{\rho}
 +\frac{4}{5} \sqrt{\frac{2}{3}} b e^{-\frac{1}{24}
\pi  \rho {\tau}^4} \sqrt{\rho} {T^0}^2 +\frac{4}{5} \sqrt{\frac{2}{3}}
b e^{-\frac{1}{24} \pi  \rho {\tau}^4} \sqrt{\rho} {\tau}^2
\label{eq:two}\\
&-\frac{b e^{-\frac{1}{24} \pi  \rho {\tau}^4} \pi  \rho^{3/2}
{T^0}^2 {\tau}^4}{5 \sqrt{6}} +\frac{b e^{-\frac{1}{24} \pi  \rho {\tau}^4} \pi  \rho^{3/2} {\tau}^6}{45 \sqrt{6}}
\label{eq:three}\\
&-\frac{1}{3}
\sqrt{\frac{2}{3}} b e^{-\frac{1}{24} \pi  \rho {\tau}^4} \pi  \rho^{3/2} {T^0} {\tau}^4 (\frac{T}{2} -T^0)-\frac{1}{3} \sqrt{\frac{2}{3}}
b e^{-\frac{1}{24} \pi  \rho {\tau}^4} \pi  \rho^{3/2} {\tau}^4 (\frac{T}{2} -T^0)^2 
\label{eq:four}\\
& +\frac{128 \sqrt{6} b {T^0}^2}{5 \pi  \sqrt{\rho} {\tau}^4} 
+\frac{112 \sqrt{6}
b e^{-\frac{1}{24} \pi  \rho {\tau}^4} {T^0}^2}{5 \pi  \sqrt{\rho} {\tau}^4}-\frac{288 b {T^0}^2 \text{Erf}[\frac{1}{2} \sqrt{\frac{\pi }{6}} \sqrt{\rho} {\tau}^2]}{\pi
 \rho {\tau}^6}
\label{eq:five}\\
& -\frac{192 \sqrt{6} b}{5 \pi  \sqrt{\rho}
{\tau}^2} +\frac{132 \sqrt{6} b e^{-\frac{1}{24} \pi  \rho {\tau}^4}}{5 \pi  \sqrt{\rho} {\tau}^2}+\frac{72 b \text{Erf}[\frac{1}{2} \sqrt{\frac{\pi }{6}} \sqrt{\rho} {\tau}^2]}{\pi  \rho {\tau}^4}\,,\label{eq:six}
\end{align}
where $\tau = \tau_{0xq}$. 

For the $x$ integral, it is again convenient to use radial null coordinates, $(u,v)$, this time centred at $q$. Then 
\begin{align}
\tau^2 &= 2 u v\,,\\
X^0 & = \frac{T}{2} - \frac{u+v}{\sqrt{2}}\,, 
\end{align}
and the integrand does not depend on the polar angles.  
The range of the integration variables is $0<v< \frac{T}{\sqrt{2}}$ and $0<u <  v$. 
As the integrand is symmetric under interchange of $u$ and $v$, the range of the $u$ integration can be extended to  $0<u <  \frac{T}{\sqrt{2}}$ if the integrand is multiplied by $\frac{1}{2}$. 

Term (\ref{eq:one}) of the integrand gives the Einstein Hilbert action. 
Of the other terms, 
the first term of (\ref{eq:two}) does not depend on $b$ and is multiplied by the conformal 
factor $(1 + 4 b (X^0)^2)$ before integrating.  Mathematica is able to analytically
integrate all the terms of the integrand except for the last three on line (\ref{eq:six}). 
Consider those three terms -- without the factor of $b$ -- as a function of $s:=\tau^2$,
\begin{align}
f_\rho(s) :=  -\frac{192 \sqrt{6} }{5 \pi  \sqrt{\rho}
s} +\frac{132 \sqrt{6}  e^{-\frac{1}{24} \pi  \rho s^2}}{5 \pi  \sqrt{\rho} {\tau}^2}+\frac{72 \text{Erf}[\frac{1}{2} \sqrt{\frac{\pi }{6}} \sqrt{\rho}s]}{\pi  \rho s^2}\,,
\end{align}
and note it is a function of $ \sqrt{\rho}s$.
\begin{figure}[h!]
\centering
{\includegraphics[scale=0.5]{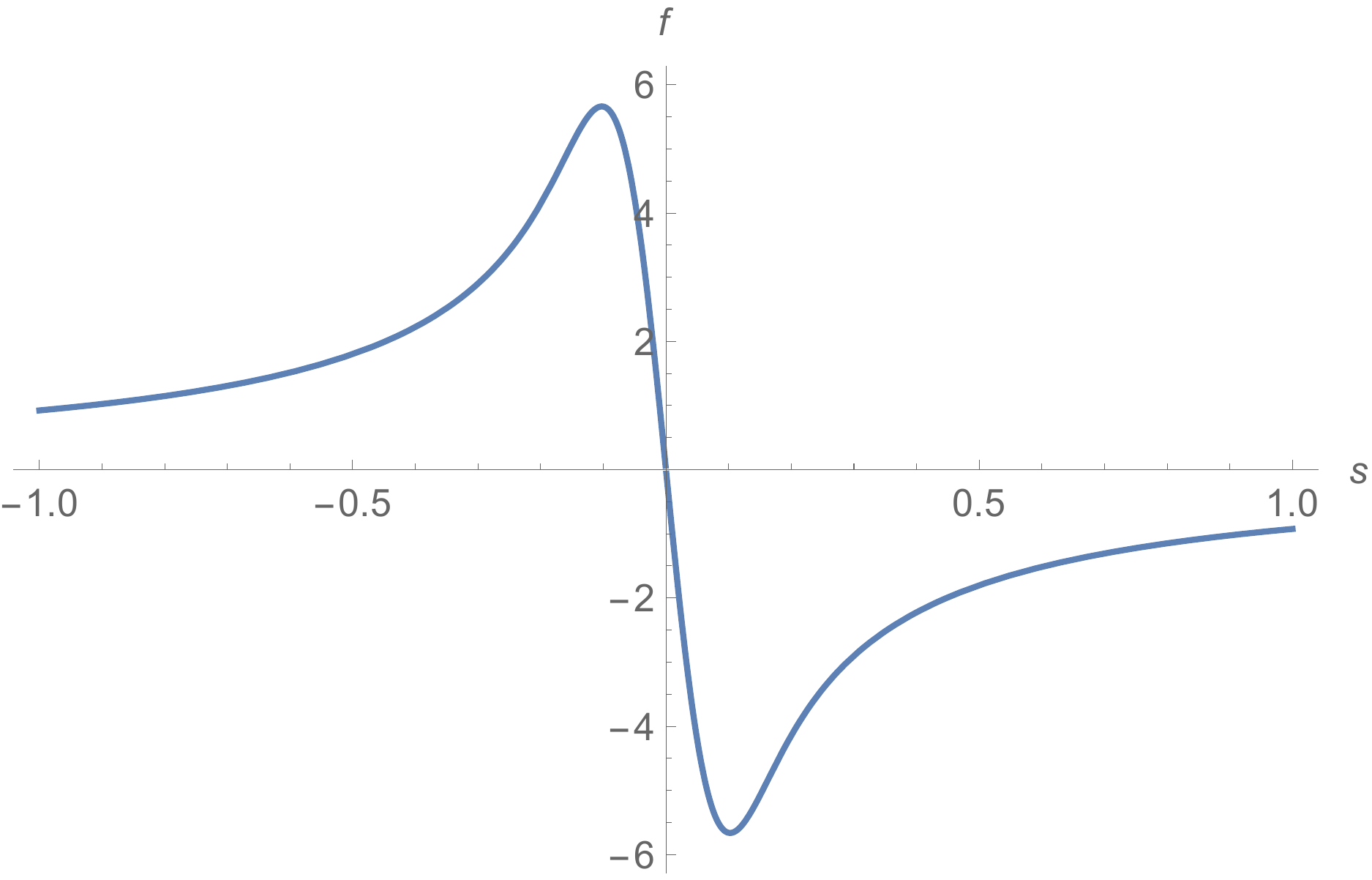}}
\caption{\label{fig:plotfun} 
$f(s)$,  where $s = \tau^2$,   for $\rho=1000$. 
}
\end{figure}
 Figure \ref{fig:plotfun} is a plot of $f(s)$ for $\rho=1000$. As $\rho$ increases the function  scales in $s$ and the peaks tend to the vertical axis without changing height, so $f(s)$ does not have a distributional character in the limit and will give a contribution of zero to the integral in the limit. 

Mathematica calculates the $x$ integral of the sum of the remaining 10 terms (not including the Einstein Hilbert term) to equal  \begin{align}
&\pi  T^2 \text{Erf}\left[\frac{1}{2} \sqrt{\frac{\pi }{6}} \sqrt{{\rho}} T^2\right] -\frac{12 b}{{\rho}}+\frac{2 \sqrt{6} e^{-\frac{1}{24} \pi  {\rho} T^4}}{\sqrt{{\rho}}}-\frac{2 \sqrt{6} {\gamma}}{\sqrt{{\rho}}}-\frac{32
\sqrt{6} b}{\pi  {\rho}^{3/2} T^2} \\
&+\frac{62 \sqrt{6} b e^{-\frac{1}{24} \pi  {\rho} T^4}}{\pi  {\rho}^{3/2} T^2}-\frac{6 \sqrt{6} b T^2}{5
\sqrt{{\rho}}}+\frac{2 \sqrt{6} b e^{-\frac{1}{24} \pi  {\rho} T^4} T^2}{\sqrt{{\rho}}}+\frac{12 b \sqrt{{\rho}} T^2}{{\rho}^{3/2}
T^2}- \frac{2\sqrt{6}}{\sqrt{\rho}}\log(\frac{\pi\rho T^4}{24})\\
&+\frac{43 b \text{Erf}\left[\frac{1}{2} \sqrt{\frac{\pi }{6}} \sqrt{{\rho}} T^2\right]}{{\rho}}-\frac{180 b \text{Erf}\left[\frac{1}{2}
\sqrt{\frac{\pi }{6}} \sqrt{{\rho}} T^2\right]}{\pi  {\rho}^2 T^4}-\frac{12 \text{Erf}\left[\frac{1}{2} \sqrt{\frac{\pi }{6}} \sqrt{{\rho}}
T^2\right]}{{\rho} T^2}\\
&-\frac{12 b \sqrt{{\rho}} T^2
\text{Erf}\left[\frac{1}{2} \sqrt{\frac{\pi }{6}} \sqrt{{\rho} T^4}\right]}{{\rho}^{3/2} T^2}+\frac{2 \sqrt{6} \text{ExpIntegralEi}\left[-\frac{1}{24}
\pi  {\rho} T^4\right]}{\sqrt{{\rho}}}\\
&+\frac{3^{3/4} b \left(\frac{2}{\pi }\right)^{1/4} \left({\rho} T^4\right)^{3/4} {\Gamma}\left[\frac{1}{4},\frac{1}{24}
\pi  {\rho} T^4\right]}{{\rho}^{3/2} T^2}-\frac{4\ 3^{3/4} b \left(\frac{2}{\pi }\right)^{1/4} \left({\rho} T^4\right)^{3/4} {\Gamma}\left[\frac{5}{4},\frac{1}{24}
\pi  {\rho} T^4\right]}{{\rho}^{3/2} T^2}\\
&-\frac{39 \sqrt{6} b T^2 \text{HypergeometricPFQ}\left[\left\{\frac{1}{2},\frac{1}{2}\right\},\left\{\frac{3}{2},\frac{3}{2}\right\},-\frac{1}{24}
\pi  {\rho} T^4\right]}{5 \sqrt{{\rho}}}\\
&+\frac{6 \sqrt{6} b T^2 \text{HypergeometricPFQ}\left[\left\{\frac{1}{2},\frac{1}{2},\frac{1}{2}\right\},\left\{\frac{3}{2},\frac{3}{2},\frac{3}{2}\right\},-\frac{1}{24}
\pi  {\rho} T^4\right]}{\sqrt{{\rho}}}
\,,
\end{align}
which tends to $\pi T^2$ in the limit.

\bibliography{../BIBLIOGRAPHY/refs}
\bibliographystyle{../BIBLIOGRAPHY/JHEP}

\end{document}